\documentclass[a4paper,fleqn,11pt]{article}
\usepackage{amsmath}
\usepackage{amssymb}
\usepackage{hyperref}
\usepackage{array}
\usepackage{calc}
\usepackage{graphicx}
\usepackage{cite}
\usepackage{mcite}
\usepackage{xspace}
\usepackage{booktabs}

\newcommand{\rbr}[1]{\left( #1\right)}
\newcommand{\abr}[1]{\langle #1\rangle}

\newcommand{\sbr}[1]{\left[ #1\right]}
\newcommand{\done}{{\rm d}}

\newcommand{\sst}{\scriptstyle}

\newcommand{\Sherpa}{S\protect\scalebox{0.8}{HERPA}\xspace} 
\newcommand{\Amegic}{A\protect\scalebox{0.8}{MEGIC++}\xspace} 
\newcommand{\Comix}{C\protect\scalebox{0.8}{OMIX}\xspace} 
 
\newcommand{\Vegas}{V\protect\scalebox{0.8}{EGAS}\xspace} 
\newcommand{\Haag}{H\protect\scalebox{0.8}{AAG}\xspace} 
\newcommand{\Csi}{C\protect\scalebox{0.8}{SI}\xspace} 
\newcolumntype{a}[2]{>{\raggedleft}p{#1}@{}p{#2}}
\newcolumntype{B}[1]{>{\centering}p{#1}}
\newcommand{\tvl}{\vrule width 1pt}
\newcommand{\thl}{\specialrule{1pt}{0pt}{0pt}}
\newcommand{\mypreprint}[1]{\begin{flushright}
  {\large #1}\end{flushright}\vspace*{2.5ex}}
\newcommand{\mytitle}[1]{{\flushleft\huge\sffamily\bfseries #1}}
\newcommand{\myauthor}[1]{{\flushleft\normalsize #1}}
\newcommand{\myinstitute}[1]{\flushleft{\small #1}\\\vspace*{5ex}}
\newcommand{\myabstract}[1]{\begin{center}
  {\large\sffamily\bfseries Abstract:
   \hspace*{1ex}}\parbox[t]{
    \textwidth*5/6-\widthof{{\large\sffamily\bfseries Abstract:}
    \hspace*{1ex}}}{#1}\end{center}}

\newcommand{\mysection}[1]{\section{\Large\sffamily\bfseries #1}}
\newcommand{\mysubsection}[1]{\subsection{\large\sffamily\bfseries #1}}
\newcommand{\mysubsubsection}[1]{\subsubsection*{\large\sffamily\bfseries #1}}

\long\def\symbolfootnote[#1]#2{\begingroup%
\def\thefootnote{\fnsymbol{footnote}}\footnote[#1]{#2}\endgroup}

\newcommand{\myfigure}[3]{
  \begin{figure}[#1]
    \begin{center}
      #2\\\myfigcaption{\widthof{#2}}{#3}
    \end{center}
  \end{figure}
}
\newcommand{\mytabcaption}[2]{
  \refstepcounter{table}
  \small{\bf Tab. \thetable\hspace*{2ex}}
  \parbox[t]{#1-\widthof{\bf Tab. \thetable\hspace*{4ex}}}{#2}
}
\newcommand{\mytable}[3]{
  \begin{table}[#1]
    \begin{center}
      #2\\\mytabcaption{\widthof{#2}}{#3}
    \end{center}
  \end{table}
}

\begin{document}
\mypreprint{IPPP/08/61\\DCPT/08/122\\[1.5mm]MCNET/08/07}
{\mytitle{How to calculate colourful cross sections efficiently}
\myauthor{Tanju Gleisberg$^1$, Stefan H{\"o}che$^2$,
  Frank Krauss$^2$, Rados{\l}aw Matyszkiewicz$^3$}
\myinstitute{$^1$ Stanford Linear Accelerator Center, 
  Stanford University, Stanford, CA 94309, USA\\
  $^2$ Institute for Particle Physics Phenomenology,
  Durham University, Durham DH1 3LE, UK\\
  $^3$ Institut f{\"u}r theoretische Physik,
  Technische Universit{\"a}t Dresden, 01062 Dresden, Germany}}

\myabstract{
Different methods for the calculation of cross sections with many QCD 
particles are compared.  To this end, CSW vertex rules, Berends-Giele 
recursion and Feynman-diagram based techniques are implemented
as well as various methods for the treatment of colours and phase 
space integration.  We find that typically there is only a small window 
of jet multiplicities, where the CSW technique has efficiencies comparable 
or better than both of the other two methods.
}

\mysection{Introduction}
\label{intro}
In the past years, a variety of string-inspired methods has been proposed 
for the efficient calculation of QCD scattering amplitudes with large 
number of external legs \cite{Cachazo:2004kj,Witten:2003nn,*Cachazo:2005ga,
  Britto:2004ap,*Britto:2005fq}.  Compared with techniques based on the 
construction of Feynman diagrams and corresponding helicity amplitudes
 \cite{Kleiss:1985yh,*Ballestrero:1994jn}, these new methods induce a 
relatively mild growth in computational complexity with increasing 
multiplicity.  A particularly interesting new technique
are the CSW vertex rules, which are based on the 
correspondence between a weakly coupled $\begin{cal}N\end{cal}=4$ Super 
Yang-Mills theory and a certain type of string theory.  It has been shown 
in \cite{Cachazo:2004kj} that this method allows to build arbitrary 
tree-level colour-ordered amplitudes from MHV amplitudes.  In the context 
of this publication, the CSW rules have been implemented into the \Sherpa 
framework \cite{Gleisberg:2003xi}. The corresponding code has been validated 
by a comprehensive comparison of cross sections with other programs for the 
processes $pp\to\; $jets and $pp\to Z+$jets.  For Monte Carlo phase space 
integration, the standard techniques presented in \cite{Byckling:1969sx} in 
the implementation of \Amegic \cite{Krauss:2001iv} are employed.  In pure 
QCD processes the \Haag generator \cite{vanHameren:2002tc} is used, which 
has been modified such that it can be employed with adaptive techniques like 
\Vegas \cite{Lepage:1980dq,*Ohl:1998jn}. 

The aim of this publication is to address the issue of efficiency of the 
CSW technique when dealing with full cross sections, including summation
over colours and helicities, rather than single amplitudes.  This extends 
previous studies presented in \cite{Dinsdale:2006sq,Duhr:2006iq}.  
For purely gluonic processes, it has already been pointed out in 
\cite{Duhr:2006iq} that Berends-Giele type recursive relations are superior 
at large final state multiplicities.  This is mainly due to the fact 
that the colour decomposition (colour ordering) of QCD amplitudes implies 
a large exponential growth in the number of required partial amplitudes, 
which can, in the case of the Berends-Giele approach, be overcome with the 
technique of colour dressing.  It was however also noted there that 
colour-ordered multi-gluon amplitudes for relatively low final state 
multiplicities ($\le 5$) are most efficiently computed using CSW vertex 
rules.  This is because in this regime the number of possible colour 
configurations is still low.  In addition, there is also only a small 
number of contributing non-MHV amplitudes, such that the major part of the
calculation involves one MHV vertex only, maximising the impact of the 
compact formulae in the CSW approach.  On the other hand, for low 
multiplicities expressions obtained from traditional Feynman-diagram based 
techniques are also compact and can easily be simplified.  The natural 
question arises, whether those methods -- Berends-Giele recursion and 
traditional Feynman-diagram based techniques -- eventually perform 
comparably or even better than the CSW vertex rules.

In this publication we aim at quantifying the effects outlined above for 
cross sections with realistic cuts and for some experimentally significant 
processes.   To this end, we compare three different numerical programs:
\begin{enumerate}
\item \Amegic \cite{Krauss:2001iv}, which computes cross sections using 
	helicity amplitudes \cite{Kleiss:1985yh,*Ballestrero:1994jn},
	constructed through Feynman diagrams;
\item the new generator \Comix \cite{comix} based on colour-dressed 
	Berends-Giele (CDBG) recursion \cite{Duhr:2006iq};
\item and a new, optimised implementation of CSW rules in the framework 
	of \Sherpa.
\end{enumerate}
Each of the methods prefers a certain type of phase space integrator.
Colour-ordered amplitudes are employed in the new implementation of the CSW 
technique and can be constructed in \Comix, such that both codes can 
employ the colour sampling integrator presented in \cite{comix}.
Since both \Amegic and \Comix have access to internal propagator lines
the techniques of \cite{Byckling:1969sx} can naturally be employed, 
since they are based on the assumed pole structure of single diagrams.  
Using \Amegic to extract corresponding integration channels from the
Feynman diagrams, this method can also be ported to be used with the CSW 
rules.

We attempt to compare as many generator-integrator pairings as possible
for a number of processes that will be relevant at the LHC.  The outline is 
as follows.  In Sec.\ \ref{sec:implementation} the algorithms needed for the 
implementation of the CSW technique are briefly summarised.  The two other 
matrix element generators have been presented in \cite{Krauss:2001iv,comix}
and we refer to the original publications for further details.  We do, 
however, comment on some relevant details of the colour treatment and the 
phase space integration, which have not been published before.  In 
Sec.\ \ref{sec:results} the programs are validated and their respective 
efficiency in terms of evaluation time per configuration and in terms of 
integration time to reach a certain precision level is discussed.

\mysection{Implementation}
\label{sec:implementation}
In this section, we briefly introduce the basic ingredients for the 
numerical implementation of the CSW rules. Phase space integration 
algorithms are presented in detail in \cite{Krauss:2001iv,comix}, 
therefore only the modified \Haag algorithm will be discussed here, cf.\ 
App.\ \ref{sec:haag}.

\mysubsection{Colour factors}
In our implementation we employ the CSW vertex rules for colour-ordered 
amplitudes; therefore the colour structure and the kinematical part of the 
amplitudes factorise.  For QCD amplitudes containing quarks the well-known 
decomposition over fundamental representation matrices $T^a_{i\bar{\jmath}}$ 
of $SU(3)$ \cite{Mangano:1990by} is used.  Considering for example one 
quark line connecting the external quarks $1$ and $n$ with colours $i_1$ and 
$\bar{\jmath}_n$ with the intermediate gluons carrying colours $a_i$, this 
decomposition reads
\begin{equation}\label{color_fund} 
  {\cal A}(1,\ldots,n) = \sum_{\sigma \in S_{n-2}}{\rm Tr}\,
    (T^{a_{\sigma_2}} \ldots T^{a_{\sigma_{n-1}}})_{i_1\bar{\jmath}_n}\; 
    A(1,\sigma_2,\ldots,\sigma_{n-1},n)\;.
\end{equation}
The sum runs over all $(n-2)!$ permutations of $(2,\ldots,n-1)$.  For purely 
gluonic amplitudes the numerically more efficient representation through 
matrices $(F^a)_{b c}=if^{abc}$, i.e.\ the adjoint representation,
 \cite{DelDuca:1999rs} is employed.  Assuming $n$ external gluons carrying
colours $a_1\ldots a_n$ this decomposition reads
\begin{equation}\label{color_ad} 
  {\cal A}(1,\ldots,n) = \sum_{\sigma \in S_{n-2}}
    (F^{a_{\sigma_2}} F^{a_{\sigma_3}} \ldots 
    F^{a_{\sigma_{n-1}}})_{a_1 a_n} \; A(1,\sigma_2,\ldots,\sigma_{n-1},n)\;,
\end{equation}
where again the sum runs over all $(n-2)!$ permutations of $(2,\ldots,n-1)$.

\mysubsection{Amplitude evaluation}
The first non-vanishing helicity configurations for tree amplitudes are the 
MHV or Parke-Taylor amplitudes \cite{Parke:1986gb,*Berends:1987me}.  They 
contain $n-2$ partons with like-sign helicity and 2 with opposite sign 
helicity.  Amplitudes with all or all but one like-sign helicity vanish at 
tree-level, cf.\ \cite{Mangano:1990by}.  The MHV ($\overline{\rm MHV}$) 
amplitudes can be written as simple holomorphic (anti-holomorphic) functions. 
In the notation of \cite{Dixon:1996wi} the $n$-gluon MHV-amplitudes read
\begin{equation}\label{mhvgluon}
  A_n\rbr{1^+,...,k^-,...,l^-,...,n^+}=
  i\,\frac{\abr{k\,l}^4}
       	    {\abr{1\,2}\abr{2\,3}\ldots\abr{n{\sst-}1\,n}\abr{n\,1}}\;.
\end{equation}
MHV amplitudes for one external quark pair are easily obtained using
supersymmetric Ward identities \cite{Mangano:1987xk,*Mangano:1987kp}. 
They are given by
\begin{equation}\label{mhvq}
  \begin{split}
  A_n\rbr{q^{+},2^+,\ldots,k^-,\ldots,n{\sst-}1^+,\bar{q}^{-}}\,=&\;
  i\,\frac{\abr{k\,q}\abr{k\,\bar q}^3}
            {\abr{1\,2}\ldots\abr{n{\sst-}1\,n}\abr{n\,1}}\\ 
  A_n\rbr{q^{-},2^+,\ldots,k^-,\ldots,n{\sst-}1^+,\bar{q}^{+}}\,=&\;
  -i\,\frac{\abr{k\,q}^3\abr{k\,\bar q}}
             {\abr{1\,2}\ldots\abr{n{\sst-}1\,n}\abr{n\,1}}
  \end{split}
\end{equation}
Corresponding $\overline{\rm MHV}$ amplitudes are obtained by complex 
conjugation, amounting to the replacement $\abr{\;}\to\sbr{\;}$.  Amplitudes 
for two external quark pairs and for the scattering into a Drell-Yan lepton 
pair have been discussed in \cite{Mangano:1988kk,*Berends:1988yn}.

The CSW vertex rules to build full amplitudes from MHV ($\overline{\rm MHV}$) 
amplitudes read \cite{Cachazo:2004kj}:
\begin{itemize}
\item For an amplitude $A_n\rbr{1^{h_1},\ldots,n^{h_n}}$ with $2<m<n-2$ 
	negative helicity external legs, draw all possible diagrams 
	connecting $m-1$ MHV ($\overline{\rm MHV}$) amplitudes. 
        Their free legs connect to the external particles.
\item Construct spinors for internal lines from the corresponding momenta 
	$p$ via $\lambda_a=p_{a\dot{b}}\eta^{\dot{b}}$,
	where $\eta$ is an arbitrary but fixed reference spinor employed in 
	all CSW diagrams.
\item Associate a scalar propagator $1/p^2$ with each internal line
	connecting two MHV vertices.
\end{itemize} 
Using this algorithm, amplitudes with an arbitrary number of negative 
(positive) helicity partons can be computed.

\mysubsection{Phase space integration using \Haag}
\label{sec:haag}
Several approaches have been presented in the past to efficiently sample
multi-particle phase spaces
\cite{Byckling:1969sx,Kleiss:1985gy,*Draggiotis:2000gm,vanHameren:2002tc}. 
For multi-jet phase space integration in pure QCD processes, the most 
advanced algorithm so far is the \Haag generator presented in 
\cite{vanHameren:2002tc}.  It is designed to produce momenta approximately 
following a QCD antenna function, cf.\ Sec.~\ref{sec:haagapp}

\mysubsubsection{Symmetrisation of antennae}
In analogy to the antenna function in Eq.~\eqref{eq:antenna}, each single 
\Haag channel can be labeled by a specific permutation of the momenta.  
Since the algorithm always starts from incoming momenta, the channels are 
invariant with respect to cyclic permutations.  However, unlike the 
underlying antenna function itself, different \Haag channels are obtained, 
if the order of the momenta is reversed. This is due to the fact that the 
open antenna algorithm is employed.  In order to recover symmetry, 
pairs of channels given by a permutation and its reverse are combined into 
one, i.e.\ one of the two configurations is chosen with equal probability 
and the weight is given by the average of the two.  Furthermore, all antennae 
can be classified into different types depending on the relative position 
of the incoming momenta ($p_0$ and $p_1$) within a permutation of the momenta.
All channels of the same type, i.e.\ with the same number of final state 
momenta between $p_0$ and $p_1$, are in principle equivalent. They can be 
obtained from each other by simply relabeling the final state momenta.

\mysubsubsection{Improvements of the algorithm}
To generate an adequate phase space integrator for realistic $n$-particle 
QCD processes, different \Haag channels can be combined using the multichannel 
method \cite{Kleiss:1994qy}, cf.\ \cite{vanHameren:2002tc}. 

The efficiency of the integrator is improved, if additionally the \Vegas 
algorithm \cite{Lepage:1980dq} is applied to optimise individual single 
channels \Vegas is very efficient in adapting to functions, whose  
peaking behaviour is not too extreme and which factorise into a product of 
one-dimensional functions.  Although this is not necessarily the case for 
a fully differential cross section, \Vegas can be used to better adapt the 
antenna-like structures in single \Haag-channels to the corresponding 
substructures in the matrix elements, including phase space cuts.

The equivalence of \Haag-channels of the same type can also be used such
that all of them employ the same \Vegas map.  This alleviates the adaptation 
significantly, since only very few maps survive, with a number that grows 
only linearly with the number of particles.  This can easily be understood
from the construction of the \Haag channels from a factorially growing number
of equivalent mappings.

\mysection{Results}
\label{sec:results}
Various approaches can be used to judge the efficiency of methods to 
evaluate amplitudes, to sum them over colours and helicities and to 
integrate them over phase space.  Finally, however it is important that the 
numerical code which implements the method yields a cross section with the
desired error as quickly as possible.  Theoretically appealing forms of the 
amplitude are not guaranteed to be of maximal help in this respect, see for 
example \cite{Duhr:2006iq}. 

Therefore, the following strategy to judge the various methods is adopted:
\begin{enumerate}
\item
The evaluation times for helicity summed amplitudes are compared.  Two 
different sets are generated, which correspond to a colour-summed and a 
colour-sampled integration, respectively.  In this publication, helicity 
sampling is not considered, since it introduces additional degrees 
of freedom which in most cases significantly slow down the integration. For 
low multiplicities, this effect is not compensated by the correspondingly 
lower number of amplitudes that have to be evaluated, cf.\ \cite{comix}.
\item
A number of cuts is imposed on the final state to yield physical cross
sections for various processes.  In each case, the integration is terminated 
once a certain precision of the result has been reached.  Note that this
effectively tests not only the method for evaluating amplitudes, but also 
the phase space sampling and optimisation.  This study corresponds exactly 
to the problem outlined above and can therefore provide vital information 
about a preference for which technique/code to use for the evaluation of 
cross sections in different regimes of particle multiplicity.
\end{enumerate}
In the following, results obtained with the new implementation of the CSW 
technique are labeled ``CSW'', the helicity methods implemented in \Amegic 
are labeled by ``AMEGIC'' and the colour-dressed Berends-Giele recursion 
implemented in \Comix is denoted by ``COMIX''. 

\mysubsection{Amplitude evaluation times}
Colour-summed amplitudes are compared in Tab.~\ref{compAME_MHV}.  The 
colour-dressed Berends-Giele recursion is not included in this comparison, since it is 
optimised to generate minimal subsets of colour-interfering amplitudes; it
would thus be very inefficient for colour-summation.  The results compiled 
here clearly show the superiority of the CSW method over the evaluation of 
Feynman diagrams for pure QCD process with no or only one quark-line.  In 
these two cases the expressions obtained from CSW rules are very compact,
leading to an increasingly better behaviour for larger numbers of final state
particles.  Additionally, the colour-decomposition in the adjoint 
representation \cite{DelDuca:1999rs} used for the CSW amplitudes seems to 
significantly simplify the computation for purely gluonic amplitudes.  For 
two quark lines the two methods are comparable in processes with lower 
multiplicities.  For higher multiplicities the standard helicity method, 
however, quickly leads to unmanageable expressions.  For processes involving 
electroweak interactions there is no obvious advantage in using CSW rules. 
 
To compare evaluation times for colour-ordered amplitudes, matrix elements
generated by traditional methods in \Amegic are not included, since the 
projection of the colour structures coming from the direct evaluation of 
full QCD Feynman rules onto colour-ordered amplitudes would not lead to a 
significant simplification compared to the full colour-summed expression.
Therefore the results for colour-ordered amplitudes shown in Tab.\
\ref{compBG_MHV} contain only CSW and Berends-Giele recursion techniques.
Obviously, the former is preferred over the latter for multiplicities that 
require the evaluation of MHV contributions only, i.e.\ processes with at 
most five particles involved.  Once next-to-MHV contributions kick in 
(six and seven particles involved) both methods exhibit a similar 
performance, beyond that the Berends-Giele technique is clearly the method 
of choice.  This corresponds to what has been found in \cite{Duhr:2006iq} 
for helicity-sampled matrix elements.  It should be noted, however, that 
when dressing amplitudes with colour and calculating corresponding 
matrix elements which include subleading colour contributions the 
colour-dressed Berends-Giele method gains additional performance compared 
to CSW rules, cf.\ \cite{Duhr:2006iq}.

\mysubsection{Integration times}	 
To investigate the integration behaviour of the various codes, the time 
needed to compute cross sections is compared for example processes at the 
LHC with a centre-of-mass energy of 14 TeV.  The setup is essentially
identical to the one employed in \cite{MC4LHC:2003aa}. Respective settings 
are listed in Tab.\ \ref{tab:mc4lhc_params}.

A variety of phase space generators is used for the integration, which all 
roughly reproduce the peaking behaviour of the matrix elements and which are 
to some extend adaptive.  For colour-summed matrix elements we use the 
improved version of the \Haag generator (``HAAG'') or an integrator using 
the standard multi-channel integration technique with channels constructed 
according to \cite{Byckling:1969sx}, (labeled ``MC'').  Their usage with 
colour-sampled matrix elements, however, turns out to be quite inefficient 
since they do not take into account that different colour assignments give 
rise to enhanced contributions in different phase space regions.  For 
colour-sampled multi-gluon scattering a special integrator has been 
constructed which is denoted by ``CSI'' (Colour Sampling Integrator).  It is 
based on the channels given by the \Haag algorithm, which are selected 
and weighted specific to a corresponding colour assignment.  For matrix 
elements generated by \Comix, a recursive phase space generator 
(denoted ``RPG'') is used which adapts the recursive structure of the 
Berends-Giele formalism and is thus applicable for very high particle 
multiplicities. The latter two integrators are described in detail in 
\cite{comix}.
 
The first class of processes is pure jet production with up to 6 jets in
the final state, for different maximal numbers of quarks.  
Tab.\ \ref{Tab:XSecsGluon} lists the results for gluon scattering processes. 
For up to four final state particles colour-summed CSW matrix elements 
combined with \Haag are most efficient.  Beyond that the number of colour 
configurations becomes large enough to render the computation of colour-summed
matrix elements a cumbersome, time-consuming exercise.  In this region CDBG
matrix elements paired with the \Csi integrator give the best performance.
In Tab. \ \ref{Tab:XSecsJets} multi-jet cross sections with one and two 
quark lines are studied, the conclusions are essentially the same as in the 
purely gluonic case.

The second class of processes is given by the production of a lepton pair 
accompanied by additional jets, with results shown in Tab.\
\ref{Tab:XSecsZJets}.   In this case, only the ``MC'' and ``RPG'' integrators 
have been employed.   Here, the best performance is achieved with 
colour-summed matrix elements for up to three jets.  As stated before the CSW 
rules do not lead to any improvement.

\section{Conclusions}
In this publication a comparison between different matrix element generators,
paired with different integration techniques has been presented. All of them 
are at present implemented in the framework of the event generator \Sherpa.

Considering the evaluation of matrix elements, it has been shown that for 
low multiplicities traditional techniques to evaluate Feynman diagrams in 
the helicity formalism perform surprisingly well.  However, with growing 
numbers of external legs, these methods quickly lead to unmanageably large 
expressions. The compact formulae for MHV amplitudes result in a significant 
gain with the CSW technique, once QCD amplitudes predominantly containing 
gluons are concerned.  Otherwise, the method performs comparably or sometimes
even worse than techniques based on Feynman diagrams.  With growing numbers 
of external legs, the colour dressed Berends-Giele recursion is the 
candidate best-suited to quickly evaluate matrix elements.

Considering the various phase space integrators, it has been shown that 
traditional methods based on the expected pole structure of the integrand 
as guessed from Feynman diagrams or similar are not only versatile enough 
to yield an appreciable convergence of results for QCD processes.  There, 
dedicated algorithms like \Haag and \Csi have better integration behaviour, 
but they are limited in their applicability.

To summarise, a wide range of matrix element evaluation and integration 
techniques has been made available within the \Sherpa framework.  The
different techniques have been tested and compared, such that for every 
given jet multiplicity an optimal performance of the overall package can 
be achieved.

\section*{Acknowledgements}
We thank Phil Roffe, Graeme Stewart and the ScotGrid \cite{ScotGrid:2008aa} 
Tier 2 sites Durham and Glasgow for technical support.   TG's research was 
supported by the US Department of Energy, contract DE-AC02-76SF00515.  SH 
thanks the HEPTOOLS Marie Curie Research Training Network (contract number 
MRTN-CT-2006-035505) for an Early Stage Researcher position.  Support from 
MCnet (contract number MRTN-CT-2006-035606) is gratefully acknowledged. 

\mytable{p}{
    \begin{tabular}{!{\tvl}p{3.5cm}!{\tvl}B{2.5cm}|B{2.5cm}|a{10mm}{12mm}@{\tvl}} 
      \thl& & & &\\[-1.5ex] Process & Time per ME & Time per ME & & \\
      & \Amegic [s] & CSW [s] & \multicolumn{2}{c!{\tvl}}{\Amegic/CSW}\\[1ex]\thl
      & & & & \\[-1.5ex] 
      $gg \to 2g$ & $6.35\times10^{-6}$ & $1.78\times10^{-6}$ & 3&.6 \\
      $gg \to 3g$ & $1.68\times10^{-4}$ & $1.73\times10^{-5}$ & 9&.7 \\
      $gg \to 4g$ & $3.63\times10^{-2}$ & $1.18\times10^{-3}$ & 31& \\
      $gg \to 5g$ & -                   & $3.29\times10^{-2}$ & &\\
      $gg \to 6g$ & -                   & $4.56$              & &\\
      $gg \to 7g$ & -                   & $280 $              & &\\[.5ex]\hline & & & & \\[-1.5ex] 
      $q\bar q \to 2g$ & $3.25\times10^{-6}$ & $1.45\times10^{-6}$ & 2&.2 \\
      $q\bar q \to 3g$ & $3.40\times10^{-5}$ & $1.21\times10^{-5}$ & 2&.8 \\ 
      $q\bar q \to 4g$ & $2.06\times10^{-3}$ & $8.80\times10^{-4}$ & 2&.3 \\
      $q\bar q \to 5g$ & $0.614$             & $2.65\times10^{-2}$ & 23& \\ 
      $q\bar q \to 6g$ & -                   & $2.96$ & &\\
      $q\bar q \to 7g$ & -                   & $170$  & &\\[.5ex] \hline & & & & \\[-1.5ex]
      $q\bar q \to q\bar q$     & $1.36\times10^{-6}$ & $2.76\times10^{-6}$ & 0&.49 \\
      $q\bar q \to q\bar q\ g$  & $1.11\times10^{-5}$ & $1.15\times10^{-5}$ & 1&.0 \\
      $q\bar q \to q\bar q\ 2g$ & $4.48\times10^{-4}$ & $5.30\times10^{-4}$ & 0&.85 \\
      $q\bar q \to q\bar q\ 3g$ & $8.98\times10^{-2}$ & $1.12\times10^{-2}$ & 8&.0 \\
      $q\bar q \to q\bar q\ 4g$ & -                   & $0.934$  & &\\
      $q\bar q \to q\bar q\ 5g$ & -                   & $42.0$ & &\\[.5ex]\hline & & & & \\[-1.5ex] 
      $q\bar q \to q^\prime\bar{q^\prime}$     & $7.99\times10^{-7}$ & $1.52\times10^{-6}$ & 0&.53 \\
      $q\bar q \to q^\prime\bar{q^\prime}\ g$  & $5.74\times10^{-6}$ & $5.65\times10^{-6}$ & 1&.0 \\
      $q\bar q \to q^\prime\bar{q^\prime}\ 2g$ & $1.07\times10^{-4}$ & $2.72\times10^{-4}$ & 0&.39 \\ 
      $q\bar q \to q^\prime\bar{q^\prime}\ 3g$ & $1.34\times10^{-2}$ & $5.80\times10^{-3}$ & 2&.3 \\
      $q\bar q \to q^\prime\bar{q^\prime}\ 4g$ & -                   & $0.470$ & &\\
      $q\bar q \to q^\prime\bar{q^\prime}\ 5g$ & -                   & $20.94$ & &\\[.5ex]\hline & & & & \\[-1.5ex] 
      $q \bar{q} \to Z(\to e^-e^+)$    & $1.56\times10^{-6}$ & $3.88\times10^{-6}$ & 0&.40 \\
      $q \bar{q} \to Z(\to e^-e^+)\ g$ & $3.99\times10^{-6}$ & $6.85\times10^{-6}$ & 0&.58 \\
      $q \bar{q} \to Z(\to e^-e^+)\ 2g$& $2.16\times10^{-5}$ & $1.07\times10^{-4}$ & 0&.20 \\ 
      $q \bar{q} \to Z(\to e^-e^+)\ 3g$& $2.34\times10^{-4}$ & $1.31\times10^{-3}$ & 0&.18 \\
      $q \bar{q} \to Z(\to e^-e^+)\ 4g$& $1.44\times10^{-2}$ & $8.20\times10^{-2}$ & 0&.18 \\[1ex]
      \thl
    \end{tabular} \vspace*{2ex}
 } {Computation times for full matrix elements summed over colour and helicity.
 Displayed are averages for a single evaluation, employing Feynman diagrams computed in the helicity formalism 
 (using \Amegic) and the Cachazo-Svr\v{c}ek-Witten (CSW) vertex rules. Numbers were generated on a 
 2.53~GHz Intel$^\text{\textregistered}$ Core\texttrademark 2 Duo T9400 CPU. \label{compAME_MHV} }
\mytable{p}{
    \begin{tabular}{!{\tvl}p{3.5cm}!{\tvl}B{2.5cm}|B{2.5cm}|a{10mm}{12mm}@{\tvl}} 
      \thl & & & & \\[-1.5ex] Process & Time per ME & Time per ME & & \\
      & BG [s] & CSW [s] & \multicolumn{2}{c!{\tvl}}{BG/CSW}\\[1ex]\thl
      & & & & \\[-1.5ex] 
      $gg \to 2g$ & $8.42\times10^{-6}$ & $1.18\times10^{-6}$ & 7&.1 \\
      $gg \to 3g$ & $3.19\times10^{-5}$ & $3.31\times10^{-6}$ & 9&.6 \\
      $gg \to 4g$ & $1.13\times10^{-4}$ & $6.09\times10^{-5}$ & 2&.0 \\
      $gg \to 5g$ & $3.58\times10^{-4}$ & $2.91\times10^{-4}$ & 1&.3 \\
      $gg \to 6g$ & $1.17\times10^{-3}$ & $6.38\times10^{-3}$ & 0&.20 \\
      $gg \to 7g$ & $3.99\times10^{-3}$ & $5.66\times10^{-2}$ & 0&.079 \\[.5ex]\hline & & & &\\[-1.5ex]
      $q\bar q \to 2g$ & $6.20\times10^{-6}$ & $1.02\times10^{-6}$ & 6&.1 \\
      $q\bar q \to 3g$ & $2.18\times10^{-5}$ & $2.46\times10^{-6}$ & 8&.9 \\
      $q\bar q \to 4g$ & $6.91\times10^{-5}$ & $4.59\times10^{-5}$ & 1&.5 \\
      $q\bar q \to 5g$ & $2.15\times10^{-4}$ & $2.34\times10^{-4}$ & 0&.92 \\
      $q\bar q \to 6g$ & $6.53\times10^{-4}$ & $4.00\times10^{-3}$ & 0&.16 \\
      $q\bar q \to 7g$ & $2.03\times10^{-3}$ & $3.11\times10^{-2}$ & 0&.065 \\[.5ex]\hline & & & &\\[-1.5ex] 
      $q\bar q \to q\bar q$     & $2.86\times10^{-6}$ & $1.56\times10^{-6}$ & 1&.8 \\
      $q\bar q \to q\bar q\ g$  & $1.17\times10^{-5}$ & $3.26\times10^{-6}$ & 3&.6 \\
      $q\bar q \to q\bar q\ 2g$ & $4.99\times10^{-5}$ & $5.92\times10^{-5}$ & 0&.84 \\
      $q\bar q \to q\bar q\ 3g$ & $1.94\times10^{-4}$ & $2.90\times10^{-4}$ & 0&.67 \\
      $q\bar q \to q\bar q\ 4g$ & $7.16\times10^{-4}$ & $4.93\times10^{-3}$ & 0&.15 \\
      $q\bar q \to q\bar q\ 5g$ & $2.86\times10^{-3}$ & $3.69\times10^{-2}$ & 0&.076 \\[.5ex]\hline & & & &\\[-1.5ex]
      $q\bar q \to q^\prime\bar{q^\prime}$     & $2.24\times10^{-6}$ & $1.06\times10^{-6}$ & 2&.1 \\
      $q\bar q \to q^\prime\bar{q^\prime}\ g$  & $8.97\times10^{-6}$ & $1.96\times10^{-6}$ & 4&.6 \\
      $q\bar q \to q^\prime\bar{q^\prime}\ 2g$ & $2.87\times10^{-5}$ & $3.39\times10^{-5}$ & 0&.85 \\
      $q\bar q \to q^\prime\bar{q^\prime}\ 3g$ & $8.18\times10^{-5}$ & $1.55\times10^{-4}$ & 0&.59 \\
      $q\bar q \to q^\prime\bar{q^\prime}\ 4g$ & $2.70\times10^{-4}$ & $2.48\times10^{-3}$ & 0&.11 \\
      $q\bar q \to q^\prime\bar{q^\prime}\ 5g$ & $8.13\times10^{-4}$ & $1.84\times10^{-2}$ & 0&.044 \\[.5ex]\hline & & & &\\[-1.5ex]
      $q \bar{q} \to Z(\to e^-e^+)$    & $3.84\times10^{-6}$ & $3.88\times10^{-6}$ & 0&.99 \\
      $q \bar{q} \to Z(\to e^-e^+)\ g$ & $1.02\times10^{-5}$ & $6.85\times10^{-6}$ & 1&.5 \\
      $q \bar{q} \to Z(\to e^-e^+)\ 2g$& $2.57\times10^{-5}$ & $6.90\times10^{-5}$ & 0&.37 \\
      $q \bar{q} \to Z(\to e^-e^+)\ 3g$& $7.06\times10^{-5}$ & $2.95\times10^{-4}$ & 0&.24 \\
      $q \bar{q} \to Z(\to e^-e^+)\ 4g$& $1.95\times10^{-4}$ & $3.72\times10^{-3}$ & 0&.052 \\[1ex]
      \thl
    \end{tabular} \vspace*{2ex} 
    }{Average computation time of partial amplitudes in multi-gluon scattering,
           summed over all helicity configurations. 
           Displayed are averages for a single evaluation, employing colour dressed 
           Berends-Giele (BG) recursion and the Cachazo-Svr\v{c}ek-Witten (CSW) vertex rules.
           Numbers were generated on a 
           2.53~GHz Intel$^\text{\textregistered}$ Core\texttrademark 2 Duo T9400 CPU.\label{compBG_MHV}}

\renewcommand\arraystretch{1.15}
\mytable{p}{
  \begin{tabular}{|c|c|}\hline
    Parameter & Value\\\hline
    \multicolumn{2}{c}{EW parameters in the $G_\mu$ scheme}\\\hline
    $G_F$ & $1.16639\times 10^{-5}$\\
    $\alpha_{QED}$ & 1/132.51\\
    $\sin^2\theta_W$ & 0.2222\\
    $M_Z$ & 91.188 GeV\\
    $m_H$ & 120 GeV\\\hline
    \multicolumn{2}{c}{QCD parameters}\\\hline
    PDF set & CTEQ6L1\\
    $\alpha_s$ & 0.130\\
    $\mu_F$, $\mu_R$ & $M_Z$\\
    jet, initial parton & $g$, $u$, $d$, $s$, $c$\\\hline
  \end{tabular}\hspace*{2ex}
  \begin{tabular}{|c|c|}\hline
    Parameter & Value\\\hline
    \multicolumn{2}{c}{Widths (fixed width scheme)}\\\hline
    $\Gamma_Z$ & 2.446 GeV\\\hline
    \multicolumn{2}{c}{Cuts}\\\hline
    $p_{\perp,\,i}$ & $> 30$ GeV\\
    $|\eta_i|$ & $< 5$\\\hline
    \multicolumn{2}{|c|}{$66\; {\rm GeV} < m_{ll} < 116\; {\rm GeV}$}\\
    \multicolumn{2}{|c|}{CDF Run~II $k_T$ algorithm \cite{Blazey:2000qt}}\\
    \multicolumn{2}{|c|}{with $k_T>30$ GeV and D=0.7}\\\hline
    \multicolumn{2}{c}{}\\
    \multicolumn{2}{c}{}\\
    \multicolumn{2}{c}{}\\
  \end{tabular}\vspace*{2ex}}{
  Parameters for the integration time comparison.\label{tab:mc4lhc_params}}
\mytable{p}{
    \begin{tabular}{!{\tvl}p{37mm}!{\tvl}B{2cm}|B{2cm}|B{2cm}|B{2cm}|B{2cm}@{\tvl}p{0mm}@{}} 
      \thl $pp\to n$ jets & &&&&&\\
      gluons only & $n=2$ & $n=3$ & $n=4$ & $n=5$ & $n=6$ &\\\hline
      MC cross section [pb]  & $8.915\cdot10^{7}$ & $5.454\cdot10^{6}$ & $1.150\cdot10^{6}$ & $2.757\cdot10^{5}$ & $7.95\cdot10^{4}$ &\\\hline
      stat. error    & 0.1\% & 0.1\% & 0.2\% & 0.5\% & 1\% &\\\hline
      & \multicolumn{5}{c@{\tvl}}{integration time for given stat. error [s] }&\\\hline
      CSW (HAAG)  & 4 & 165 & 1681 & 12800 & $2\cdot10^{6}$ &\\\hline
      CSW (CSI)   & - & 480 & 6500 & 11900 & 197000 &\\\hline
      AMEGIC (HAAG) & 6 & 492 & 41400 &-&-&\\\hline
      COMIX (RPG)  & 159 & 5050 & 33000 & 38000 & 74000 &\\\hline
      COMIX (CSI)  & - & 780 & 6930 & 6800 & 12400 &\\\thl
    \end{tabular}\vspace*{1ex}
  }{Cross section and evaluation times for different matrix element (phase space) generation methods
    for multi-gluon scattering at the LHC, given in pb. Numbers were generated on a 
 2.53~GHz Intel$^\text{\textregistered}$ Core\texttrademark 2 Duo T9400 CPU.
    For cuts and parameter settings, cf.\ Tab.~\ref{tab:mc4lhc_params}.
  \label{Tab:XSecsGluon}}

\begin{table}[p]
  \begin{center}
    \begin{tabular}{!{\tvl}p{37mm}!{\tvl}B{2cm}|B{2cm}|B{2cm}|B{2cm}|B{2cm}@{\tvl}p{0mm}@{}} 
      \thl $pp\to n$ jets & &&&&&\\
      $\le$ 1 quark line & $n=2$ & $n=3$ & $n=4$ & $n=5$ & $n=6$ &\\\hline
      MC cross section [pb]  & $1.456\cdot10^{8}$ & $1.051\cdot10^{7}$ & $2.490\cdot10^{6}$ & $6.75\cdot10^{5}$ & $2.14\cdot10^{5}$ &\\\hline
      stat. error    & 0.1\% & 0.1\% & 0.2\% & 0.5\% & 1\% &\\\hline
      & \multicolumn{5}{c@{\tvl}}{integration time for given stat. error [s] }&\\\hline
      CSW (HAAG)  & 10 & 354 & 6980 & 60000 & $9\cdot10^{6}$ &\\\hline
      AMEGIC (HAAG) & 13 & 930 & 73000 &-&-&\\\hline
      COMIX (RPG)  & 254 & 5370 & 15900 & 36800 & 64100 &\\\thl
    \end{tabular}\vspace*{1ex}
    \begin{tabular}{!{\tvl}p{37mm}!{\tvl}B{2cm}|B{2cm}|B{2cm}|B{2cm}|B{2cm}@{\tvl}p{0mm}@{}} 
      \thl $pp\to n$ jets & &&&&&\\
      $\le$ 2 quark lines & $n=2$ & $n=3$ & $n=4$ & $n=5$ & $n=6$ &\\\hline
      MC cross section [pb]  & $1.5129\cdot10^{8}$ & $1.1198\cdot10^{7}$ & $2.831\cdot10^{6}$ & $8.12\cdot10^{5}$ & $2.71\cdot10^{5}$ &\\\hline
      stat. error    & 0.1\% & 0.1\% & 0.2\% & 0.5\% & 1\% &\\\hline
      & \multicolumn{5}{c@{\tvl}}{integration time for given stat. error [s] }&\\\hline
      CSW (HAAG)  & 16 & 730  & 12300 & 120000 & $2\cdot10^{7}$ &\\\hline
      AMEGIC (HAAG) & 19 & 1530 & 78000 &-&-&\\\hline
      COMIX (RPG)  & 525 & 10800 & 25600 & 59000 & 113000 &\\\thl
    \end{tabular}\\[1ex]
  \mytabcaption{0.99\textwidth}{Cross section and evaluation times for different matrix element (phase space) generation methods
     for multi-jet production at the LHC, given in pb. Numbers were generated on a 
 2.53~GHz Intel$^\text{\textregistered}$ Core\texttrademark 2 Duo T9400 CPU.
     For cuts and parameter settings, cf.\ Tab.~\ref{tab:mc4lhc_params}.
  \label{Tab:XSecsJets}}
  \end{center}
\end{table}
\mytable{p}{
    \begin{tabular}{!{\tvl}p{37mm}!{\tvl}B{2cm}|B{2cm}|B{2cm}|B{2cm}|B{2cm}@{\tvl}p{0mm}@{}} 
      \thl $pp\to Z+$ jets & $n=0$ & $n=1$ & $n=2$ & $n=3$ & $n=4$ &\\\hline
       MC cross section [pb]  & 1080.8 & 121.67 & 54.67 & 23.59 & 11.22 &\\\hline
      stat. error    & 0.1\% & 0.1\% & 0.1\% & 0.2\% & 0.5\% &\\\hline
      & \multicolumn{5}{c@{\tvl}}{integration time for given stat. error [s] }&\\\hline
      CSW (MC) & 12 & 210 & 4100 & 57000 & 1500000 &\\\hline
      AMEGIC (MC) & 7 & 98 & 1060 & 10400 & 310000 &\\\hline
      COMIX (RPG)  & 15 & 364 & 6400 & 16400 & 54000 &\\\thl
    \end{tabular}\vspace*{1ex}
  }{Cross section and evaluation times for different matrix element (phase space) generation methods
    for Z+jet production at the LHC, given in pb. Numbers were generated on a 
 2.53~GHz Intel$^\text{\textregistered}$ Core\texttrademark 2 Duo T9400 CPU.
    For cuts and parameter settings, cf.\ Tab.~\ref{tab:mc4lhc_params}.
  \label{Tab:XSecsZJets}}

\appendix
\mysection{The \Vegas-improved \Haag algorithm}
\label{sec:haagapp}
The \Haag phase space generator~\cite{vanHameren:2002tc} is designed to 
produce momenta distributed approximately according to a QCD 
antenna function for an $n$-particle process
\begin{equation}\label{eq:antenna}
  A_{n}(p_0,p_1,...,p_{n-1})=\frac{1}{
    (p_0 p_1)(p_1 p_2)...(p_{n-2} p_{n-1})(p_{n-1} p_0)}.
\end{equation}
Different antennae can be obtained from permutations of the momenta $p_i$.
Cyclic permutation and reversion of the order will however lead to
the same structure. Generally \Haag relies on phase space factorisation 
over time-like intermediate momenta.
In Ref.~\cite{vanHameren:2002tc} two algorithms are proposed which are referred to 
as closed and open antenna and which differ in the decomposition of the 2-particle
phase space $d\Phi_2$. Only the closed antenna contains all factors in
Eq.~\eqref{eq:antenna}, while in the open antenna one factor $(p_i p_{i+1})$
is missing. Although the closed antenna seems to be more symmetric, in practice
it turns out that the open antenna is more efficient. This is mainly due
to its simpler structure and less additional weight factors that appear within the 
algorithm.\footnote{ These weights are nonsingular in any of the products $(p_i p_j)$.}
In the following we will therefore focus on open antennas. The algorithm is 
reviewed for the case of massless external particles, however it can
easily be generalised to the massive case.

\mysubsection{Antenna Generation}
\label{sec:haag_channels}
In the following we use a classification of antenna types by the position 
of the incoming momenta, $p_0$ and $p_1$, within the antenna, 
see Fig.~\ref{fig:atypes}. The type is then given by ${\rm Min}(m-1,n-m-1)$.

The basic building block for antenna generation is the split of a 
massive momentum according to the phase space element 
$\done s\,\done\Phi_2(Q;p,P;q)$, where $P^2=s$ and the last argument, 
$q$, defines an axis for the momentum generation. We further decompose
\begin{equation}
  \done\Phi_2(Q;p,P;q)=\done a\,\done\phi\;,\quad\text{where}\quad
  a=\frac{q\cdot p}{q\cdot P}
\end{equation}
and $\phi$ is an azimuthal angle around $q$.

The phase space for a single split, now defined through the variables 
$s,\,a,\,\phi$, is constructed as follows\footnote{ Frame dependent quantities 
are defined in the CM frame of $Q$ with the $z$-Axis along $q$}:
\begin{enumerate}
  \item Dice $s$ according to $1/s$ in $[s_{\rm min},s_{\rm max}]$.
  \item Dice $a$ according to $1/a$ in $[a_{\rm min},a_{\rm max}]$.
  \item Dice $\phi$ according to a flat distribution in $[0,2\pi]$.
  \item The momenta are given by
  \begin{equation}
    \begin{split}
      p=&\left(\frac{Q^2-s}{2\sqrt{Q^2}},\vec p\right)\;, \\
      P=&\left(\frac{Q^2+s}{2\sqrt{Q^2}},-\vec p\right)\;, \\
      \vec p=&\left(h\cos{\phi},h\sin{\phi},\frac{Q^2(1-2a)-s}{2\sqrt{Q^2}}\right)\;,
      \quad\text{where}\quad h=\sqrt{Q^2a(1-a)-as}.
    \end{split}
  \end{equation}
  \item The weight is given by
  \begin{equation}
    \frac{g(s_{\rm min},s_{\rm max})}s \frac{g(a_{\rm min},a_{\rm max})}a \frac1{2\pi}\;,
    \quad\text{where}\quad
    g(x_{\rm min},x_{\rm max})=\log{\frac{x_{\rm max}}{x_{\rm min}}}\,. 
  \end{equation}
\end{enumerate}

\mysubsubsection{Type 0 antennae}
The phase space for type 0 antenna configurations can be obtained by a direct 
multiple application of the basic building block:
\begin{equation}
  \begin{split}
    \done\Phi_n(p_0,p_1;p_2,...,p_{n-1})=
      \hphantom{\times}&\;\done s_2\;\done\Phi_2(p_0+p_1;p_2,Q_2;p_1) \\
      \times&\;\done s_3\;\done\Phi_2(Q_2;p_3,Q_3;p_2) \\
      &\quad\vdots\\
      \times&\;\done s_{n-3}\;\done\Phi_2(Q_{n-4};p_{n-3},Q_{n-3};p_{n-4}) \\
      \times&\;\done\Phi_2(Q_{n-3};p_{n-1},p_{n-2};p_{n-3})\;.
  \end{split}
\end{equation}
The corresponding total weight is given by
\begin{equation}
  w\sim\frac{\prod_{j=1}^{n-3}p_j\left(\sum_{i=j+1}^{n-1}p_i\right)}
    {\prod_{j=3}^{n-3}\left(\sum_{i=j}^{n-1}p_i\right)^2}\,
    \frac1{(p_1\!\cdot\!p_2)(p_2\!\cdot\!p_3)\cdots(p_{n-2}\!\cdot\!p_{n-1})}\,,
\end{equation}
where the contributions from boundary dependent functions $g$ have been omitted.
\myfigure{t}{
  \begin{minipage}{0.39\textwidth}
    \includegraphics[width=\textwidth]{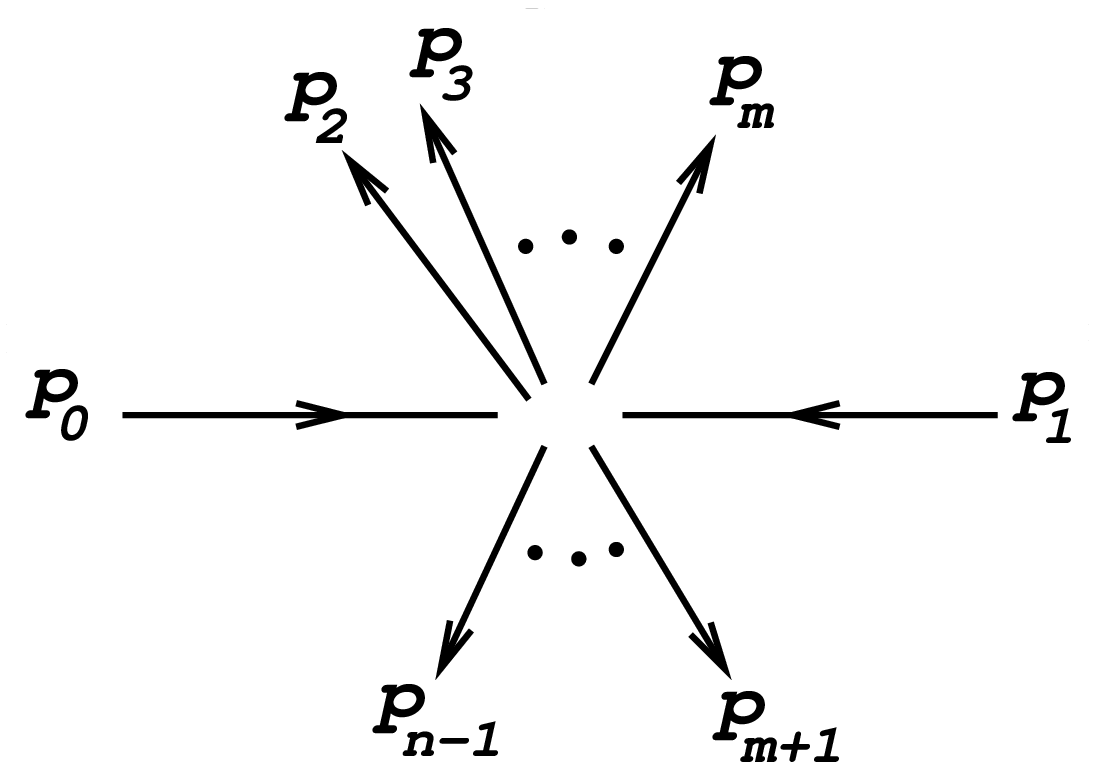}
  \end{minipage}
  \begin{minipage}{0.59\textwidth}
    $\sim\displaystyle\frac1{(p_0\!\cdot\! p_2)(p_2\!\cdot\! p_3)...
      (p_m\!\cdot\! p_1)(p_1\!\cdot\! p_{m+1})...
      (p_{n-2}\!\cdot\! p_{n-1})(p_{n-1}\!\cdot\! p_0)}$
  \end{minipage}\vspace*{5mm}}{
  Antenna configuration.\label{fig:atypes}}

\mysubsubsection{Type 1 antennae}
For this configuration the following phase space decomposition is considered:
\begin{equation}
  \begin{split}
    \done\Phi_n(p_0,p_1;p_2,...,p_{n-1})=
      \hphantom{\times}&\;\done s_2\;\done\Phi_2(p_0+p_1;p_2,Q_2;p_0) \\
      \times&\;\done s_3\;\done\Phi_2(Q_2;p_3,Q_3;p_1) \\
      \times&\;\done s_4\;\done\Phi_2(Q_3;p_4,Q_4;p_3) \\
      &\quad\vdots\\
      \times&\;\done s_{n-3}\;\done\Phi_2(Q_{n-4};p_{n-3},Q_{n-3};p_{n-4}) \\
      \times&\;\done\Phi_2(Q_{n-3};p_{n-1},p_{n-2};p_{n-3})\;.
  \end{split}
\end{equation}
In the first momentum split, $\done\Phi_2(p_0+p_1;p_2,Q_2;p_0)$,
the variable $a$ is now diced according to $\frac{1}{a(1-a)}$.
All following splits are generated according to the basic building block.
The corresponding total weight is given by
\begin{equation}
w\sim p_0\!\cdot\!\!(p_0+p_1-p_2)\,p_1\!\!\cdot\!\!(p_0+p_1-p_2)
\frac{\prod_{j=3}^{n-3}p_j\left(\sum_{i=j+1}^{n-1}p_i\right)}
{\prod_{j=3}^{n-3}\left(\sum_{i=j}^{n-1}p_i\right)^2}\,
\frac1{(p_0\!\cdot\!p_2)(p_1\!\cdot\!p_3)\cdots(p_{n-2}\!\cdot\!p_{n-1})}\,.
\end{equation}

\mysubsubsection{Type $k$ ($\geq 2$) antennae}
In this case we have the following decomposition:
\begin{equation}
  \begin{split}
    \done\Phi_n(p_0,p_1;p_2,...,p_{n-1})=
      \hphantom{\times}&\;\done s_2\done s_k\;\done\Phi_2(p_0+p_1;p_2,Q_k;p_0)\\
      \times&\;\done s_3\;\done\Phi_2(Q_2;p_3,Q_3;p_0)\\
      &\quad\vdots\\
      \times&\;\done s_{k-2}\;\done\Phi_2(Q_{k-3};p_{k-2},Q_{k-2};p_{k-3})\\
      \times&\;\done\Phi_2(Q_{k-2};p_{k-1},p_{k};p_{k-2})\\
      \times&\;\done s_{k+1}\;\done\Phi_2(Q_k;p_{k+1},Q_{k+1};p_1)\\
      &\quad\vdots\\
      \times&\;\done s_{n-3}\;\done\Phi_2(Q_{n-4};p_{n-3},Q_{n-3};p_{n-4})\\
      \times&\;\done\Phi_2(Q_{n-3};p_{n-1},p_{n-2};p_{n-3})\;.
  \end{split}
\end{equation}
All splittings are generated according to the basic building block.
The corresponding total weight is given by
\begin{equation}
  \begin{split}
    w\sim\hphantom{\times}&\; p_1\!\cdot\!\!(p_{k+1}+...+p_{n-1})
    \frac{\prod_{j=2}^{k-2}p_j\left(\sum_{i=j+1}^{k}p_i\right)}
    {\prod_{j=2}^{k-2}\left(\sum_{i=j}^{k}p_i\right)^2}\,
    \frac{\prod_{j=k+1}^{n-3}p_j\left(\sum_{i=j+1}^{n-1}p_i\right)}
    {\prod_{j=k+1}^{n-3}\left(\sum_{i=j}^{n-1}p_i\right)^2}\nonumber \\
    \times&\;\frac1{(p_0\!\cdot\!p_2)(p_2\!\cdot\!p_3)\cdots(p_{k-1}\!
      \cdot\!p_k)(p_1\!\cdot\!p_{k+1})
    \cdots(p_{n-2}\!\cdot\!p_{n-1})}\,.
  \end{split}
\end{equation}

\mysubsection{\Haag and variance reducing techniques}
To generate an adequate phase space integrator for realistic 
$n$-particle QCD processes, different \Haag channels can be combined 
using the multi-channel method~\cite{Kleiss:1994qy}. 
Symbolically we can write a single channel as a map $X$ from uniformly 
distributed random numbers $\vec a\in\sbr{0,1}^{3n-4}$
to the four-momenta $\vec p=\rbr{p_1,\ldots,p_n}$ of external particles,
The corresponding phase space weight $g$ is given by
\begin{equation}
  \frac{1}{g}=\frac{\done\Phi_n(X(\vec a))}{\done\vec a}\,.
\end{equation}
The multi-channel method now combines several maps $X_i$ to a new map
as follows:
\begin{equation}\label{eq:gen_point_mc}
  {\bf X}(\vec a,\tilde\alpha)=X_k(\vec a)\;,\quad\text{for}\quad
  \sum_{l=1}^{k-1}\alpha_l<\tilde\alpha<\sum_{l=1}^{k}\alpha_l\;,
\end{equation}
requiring an additional random number $\tilde\alpha$ and arbitrary
coefficients $\alpha_k$ with $\alpha_k>0$ and $\sum_k \alpha_k=1$. 
The corresponding phase space weight is given by
\begin{equation}\label{eq:gen_weight_mc}
  G=\sum_k \alpha_k\; g_k\;.
\end{equation}
The coefficients $\alpha_k$ can be adapted such that the variance of the 
phase space integral is minimised.

The efficiency of the integrator is improved if additionally the \Vegas 
algorithm~\cite{Lepage:1980dq} is applied to the single channels. 
\Vegas is very efficient in the numerical adaptation to functions, 
whose peaking behaviour is not too extreme and which are factorisable
to a product of one-dimensional functions. Although this is usually
not the case for a full differential cross section, it can be used 
to better adapt the antenna-like structures in a single \Haag-channel
to the corresponding structures in the matrix elements, including 
phase space cuts.

For each channel $k$, \Vegas is used to generate a mapping $\xi_k$ from 
uniformly distributed random numbers to a non-uniform distribution, 
still inside the interval $[0,1]$, and a corresponding weight $v_k$.  
To combine this with the multi-channel method the mapping $X(\vec a)$ 
for single channels must be invertible, which is the case for 
\Haag channels. The full map reads
\begin{eqnarray}
{\bf X}(\vec a,\tilde\alpha)&=&X_k(\xi_k(\vec a))\;,\;\;{\rm for}\;\;
\sum_{l=1}^{k-1}\alpha_l<\tilde\alpha<\sum_{l=1}^{k}\alpha_l\;.
\end{eqnarray}
For a momentum configuration $\vec p$ the weight is therefore given by
\begin{equation}
G(\vec p)=\sum_k \alpha_k\; g_k(\vec p)\; v_k(X_k^{-1}(\vec p))\;.
\end{equation}
We can make use of the equivalence of \Haag-channels of the same type,
such that all of them employ the same \Vegas map. This alleviates 
the adaptation significantly, since we are left with only very few maps 
and a linear growth with the number of particles.

\bibliographystyle{amsunsrt}  
\bibliography{bibliography}

\begin{thebibliography}{10}
\addtolength{\itemsep}{1ex-\baselineskip}

\bibitem{Cachazo:2004kj}
F.~Cachazo, P.~Svr\v{c}ek and E.~Witten, \emph{MHV vertices and tree amplitudes
  in gauge theory}, JHEP \textbf{09} (2004), 006,
  [\href{http://xxx.lanl.gov/abs/hep-th/0403047}{{\tt hep-th/0403047}}]. \relax
 \relax
\bibitem{Witten:2003nn}
E.~Witten, \emph{Perturbative Gauge Theory as a String Theory in Twistor
  Space}, Commun. Math. Phys. \textbf{252} (2004), 189--258,
  [\href{http://xxx.lanl.gov/abs/hep-th/0312171}{{\tt hep-th/0312171}}]. \relax
 \relax
\bibitem{Cachazo:2005ga}
F.~Cachazo and P.~Svr{\v{c}}ek, \emph{Lectures on Twistor Strings and
  Perturbative Yang-Mills Theory}, PoS \textbf{RTN2005} (2005), 004,
  [\href{http://xxx.lanl.gov/abs/hep-th/0504194}{{\tt hep-th/0504194}}]. \relax
 \relax
\bibitem{Britto:2004ap}
R.~Britto, F.~Cachazo and B.~Feng, \emph{New Recursion Relations for Tree
  Amplitudes of Gluons}, Nucl. Phys. \textbf{B715} (2005), 499--522,
  [\href{http://xxx.lanl.gov/abs/hep-th/0412308}{{\tt hep-th/0412308}}]. \relax
 \relax
\bibitem{Britto:2005fq}
R.~Britto, F.~Cachazo, B.~Feng and E.~Witten, \emph{Direct proof of tree-level
  recursion relation in Yang-Mills theory}, Phys. Rev. Lett. \textbf{94}
  (2005), 181602,  [\href{http://xxx.lanl.gov/abs/hep-th/0501052}{{\tt
  hep-th/0501052}}]. \relax
 \relax
\bibitem{Kleiss:1985yh}
R.~Kleiss and W.~J. Stirling, \emph{{Spinor techniques for calculating
  $p\bar{p}\to W^\pm/Z^0$ + jets}}, Nucl. Phys. \textbf{B262} (1985), 235--262.
  \relax
 \relax
\bibitem{Ballestrero:1994jn}
A.~Ballestrero and E.~Maina, \emph{{A new method for helicity calculations}},
  Phys. Lett. \textbf{B350} (1995), 225--233,
  [\href{http://xxx.lanl.gov/abs/hep-ph/9403244}{{\tt hep-ph/9403244}}]. \relax
 \relax
\bibitem{Gleisberg:2003xi}
T.~Gleisberg, S.~H{\"o}che, F.~Krauss, A.~Sch{\"a}licke, S.~Schumann and
  J.~Winter, \emph{\Sherpa 1.$\alpha$, a proof-of-concept version}, JHEP
  \textbf{02} (2004), 056,  [\href{http://xxx.lanl.gov/abs/hep-ph/0311263}{{\tt
  hep-ph/0311263}}]. \relax
 \relax
\bibitem{Byckling:1969sx}
E.~Byckling and K.~Kajantie, \emph{N-particle phase space in terms of invariant
  momentum transfers}, Nucl. Phys. \textbf{B9} (1969), 568--576. \relax
 \relax
\bibitem{Krauss:2001iv}
F.~Krauss, R.~Kuhn and G.~Soff, \emph{AMEGIC++ 1.0: A Matrix Element Generator
  In C++}, JHEP \textbf{02} (2002), 044,
  [\href{http://xxx.lanl.gov/abs/hep-ph/0109036}{{\tt hep-ph/0109036}}]. \relax
 \relax
\bibitem{vanHameren:2002tc}
A.~van Hameren and C.~G. Papadopoulos, \emph{A hierarchical phase space
  generator for QCD antenna structures}, Eur. Phys. J. \textbf{C25} (2002),
  563--574,  [\href{http://xxx.lanl.gov/abs/hep-ph/0204055}{{\tt
  hep-ph/0204055}}]. \relax
 \relax
\bibitem{Lepage:1980dq}
G.~P. Lepage, \emph{VEGAS: An Adaptive Multi-dimensional Integration Program},
  CLNS-80/447. \relax
 \relax
\bibitem{Ohl:1998jn}
T.~Ohl, \emph{Vegas revisited: Adaptive Monte Carlo integration beyond
  factorization}, Comput. Phys. Commun. \textbf{120} (1999), 13--19,
  [\href{http://xxx.lanl.gov/abs/hep-ph/9806432}{{\tt hep-ph/9806432}}]. \relax
 \relax
\bibitem{Dinsdale:2006sq}
M.~Dinsdale, M.~Ternick and S.~Weinzierl, \emph{A comparison of efficient
  methods for the computation of Born gluon amplitudes}, JHEP \textbf{03}
  (2006), 056,  [\href{http://xxx.lanl.gov/abs/hep-ph/0602204}{{\tt
  hep-ph/0602204}}]. \relax
 \relax
\bibitem{Duhr:2006iq}
C.~Duhr, S.~H{\"o}che and F.~Maltoni, \emph{Color-dressed recursive relations
  for multi-parton amplitudes}, JHEP \textbf{08} (2006), 062,
  [\href{http://xxx.lanl.gov/abs/hep-ph/0607057}{{\tt hep-ph/0607057}}]. \relax
 \relax
\bibitem{comix}
T.~Gleisberg and S.~H{\"o}che, \emph{Comix, a new Matrix Element generator},
  \href{http://xxx.lanl.gov/abs/0808.3674}{{\tt arXiv:0808.3674 [hep-ph]}}.
  \relax
 \relax
\bibitem{Mangano:1990by}
M.~L. Mangano and S.~J. Parke, \emph{Multi-Parton Amplitudes in Gauge
  Theories}, Phys. Rept. \textbf{200} (1991), 301--367,
  [\href{http://xxx.lanl.gov/abs/hep-th/0509223}{{\tt hep-th/0509223}}]. \relax
 \relax
\bibitem{DelDuca:1999rs}
V.~Del~Duca, L.~J. Dixon and F.~Maltoni, \emph{New Color Decompositions for
  Gauge Amplitudes at Tree and Loop Level}, Nucl. Phys. \textbf{B571} (2000),
  51--70,  [\href{http://xxx.lanl.gov/abs/hep-ph/9910563}{{\tt
  hep-ph/9910563}}]. \relax
 \relax
\bibitem{Parke:1986gb}
S.~J. Parke and T.~R. Taylor, \emph{Amplitude for n-Gluon Scattering}, Phys.
  Rev. Lett. \textbf{56} (1986), 2459. \relax
 \relax
\bibitem{Berends:1987me}
F.~A. Berends and W.~T. Giele, \emph{Recursive calculations for processes with
  n gluons}, Nucl. Phys. \textbf{B306} (1988), 759. \relax
 \relax
\bibitem{Dixon:1996wi}
L.~J. Dixon, \emph{Calculating scattering amplitudes efficiently},
  \href{http://xxx.lanl.gov/abs/hep-ph/9601359}{{\tt hep-ph/9601359}}. \relax
 \relax
\bibitem{Mangano:1987xk}
M.~L. Mangano, S.~J. Parke and Z.~Xu, \emph{Duality and multi-gluon
  scattering}, Nucl. Phys. \textbf{B298} (1988), 653. \relax
 \relax
\bibitem{Mangano:1987kp}
M.~L. Mangano and S.~J. Parke, \emph{{Quark - gluon amplitudes in the dual
  expansion}}, Nucl. Phys. \textbf{B299} (1988), 673. \relax
 \relax
\bibitem{Mangano:1988kk}
M.~L. Mangano, \emph{{The color Structure of gluon emission}}, Nucl. Phys.
  \textbf{B309} (1988), 461. \relax
 \relax
\bibitem{Berends:1988yn}
F.~A. Berends, W.~T. Giele and H.~Kuijf, \emph{{Exact expressions for processes
  involving a vector boson and up to five partons}}, Nucl. Phys. \textbf{B321}
  (1989), 39. \relax
 \relax
\bibitem{Kleiss:1985gy}
R.~Kleiss, W.~J. Stirling and S.~D. Ellis, \emph{{A new Monte Carlo treatment
  of multiparticle phase space at high energies}}, Comput. Phys. Commun.
  \textbf{40} (1986), 359. \relax
 \relax
\bibitem{Draggiotis:2000gm}
P.~D. Draggiotis, A.~van Hameren and R.~Kleiss, \emph{{SARGE: an algorithm for
  generating QCD-antennas}}, Phys. Lett. \textbf{B483} (2000), 124--130,
  [\href{http://xxx.lanl.gov/abs/hep-ph/0004047}{{\tt hep-ph/0004047}}]. \relax
 \relax
\bibitem{Kleiss:1994qy}
R.~Kleiss and R.~Pittau, \emph{Weight optimization in multichannel Monte
  Carlo}, Comput. Phys. Commun. \textbf{83} (1994), 141--146,
  [\href{http://xxx.lanl.gov/abs/hep-ph/9405257}{{\tt hep-ph/9405257}}]. \relax
 \relax
\bibitem{MC4LHC:2003aa}
 \href{http://mlm.home.cern.ch/mlm/mcwshop03/mcwshop.html}{{\tt
  http://mlm.home.cern.ch/mlm/mcwshop03/mcwshop.html}}. \relax
 \relax
\bibitem{ScotGrid:2008aa}
 \href{http://www.scotgrid.ac.uk}{{\tt http://www.scotgrid.ac.uk}}. \relax
 \relax
\bibitem{Blazey:2000qt}
G.~C. Blazey et~al., \emph{{Run II Jet Physics}},
  \href{http://xxx.lanl.gov/abs/hep-ex/0005012}{{\tt hep-ex/0005012}}. \relax
 \relax
\end{thebibliography}
\end{document}